\documentstyle[psfig,twocolumn,prl,aps,amssymb]{revtex}
\begin{document}
\wideabs{
\draft

\date{\today} 
\title{Relevance of inter-composite fermion interaction to the edge Tomonaga-Luttinger liquid}
\author{Sudhansu S. Mandal and Jainendra K. Jain}
\address{Department of Physics, 104 Davey Laboratory, The Pennsylvania State 
University, University Park, Pennsylvania 16802}
\maketitle

\begin{abstract}

It is shown that Wen's effective theory correctly describes the Tomonaga-Luttinger liquid at 
the edge of a system of {\em non}-interacting composite fermions.  However, 
the weak residual interaction between composite fermions appears to be a relevant 
perturbation.  The filling factor dependence of the Tomonaga-Luttinger parameter is estimated for 
interacting composite fermions in a microscopic 
approach and satisfactory agreement with experiment is achieved.
It is suggested that the electron field operator may not have a simple representation in the 
effective one dimensional theory.

\end{abstract}

}

A central assertion of Wen's effective theory of the edge liquid in the fractional 
quantum Hall effect (FQHE) is  
that its Tomonaga-Luttinger (TL) exponent, which describes the long distance behavior of 
various correlation functions, is a topological quantum number characteristic of the FQHE state 
in the bulk, insensitive to perturbations that do not affect the Hall quantization \cite{Wen}.  
In particular, for fractions $\nu=n/(2n+1)$ its value is predicted to be $\alpha=3$.
This would imply that the edge states in the FQHE constitute an example of a ``universal" 
Tomonaga-Luttinger liquid, in contrast to the usual
TL liquids for which the exponent varies continuously with the strength of the interaction.
However, this result is not derivable rigorously
from first principles, and therefore it is important to subject it to independent tests.

The tunneling experiment of Chang {\em et al.}\cite{Chang1} nicely demonstrated that the 
FQHE edges form a TL liquid; a power-law behavior is observed over many decades in 
the I-V characteristics, from which the edge exponent may be determined.  Several recent 
experiments \cite{Grayson,Chang3,Hilke} have studied the filling factor dependence in detail.
They find that the edge exponent varies smoothly along the sequence 
$\nu=n/(2n+1)$, does not exhibit well quantized plateaus concurrent with the FQHE plateaus in 
resistance\cite{shoulder}, and is 
sample dependent.  These experiments have motivated a number of theoretical studies 
\cite{Mandal,Conti,Shytov,MacDonald,Fradkin,Goldman}.

We have investigated this issue in a microscopic approach.  Our principal findings, discussed below 
in more detail, are as follows.  (i) A study of several filling factors of the form $n/(2n+1)$ 
suggests the remarkable  
result that the edge exponent is $\alpha=3$ for {\em non}-interacting composite 
fermions (CF's), but changes from this value when the interaction between composite fermions is taken 
into account. Even though the CF-CF interaction is weak, the corrections to $\alpha$ can be substantial.
(ii) We have estimated the exponent for the Coulomb interaction and found 
its filling factor dependence to be in satisfactory agreement with that seen experimentally.  
(iii) We argue that for interacting composite fermions, the electron field operator 
may not have a simple form in the effective one dimensional theory, and speculate on how 
that might alter the exponent.

Following Refs.~\cite{Mandal,Lee}, 
we will deduce the edge exponent from the equal time edge Green's function, defined as
\begin{equation}
G_{edge}({\bf r}-{\bf r}')={<\psi|\Psi_e^\dagger({\bf r})\Psi_e({\bf r}')|\psi>\over <\psi|\psi>}
\end{equation}
where $\psi$ is the ground state, $\Psi_e$ and $\Psi_e^\dagger$ are annihilation and creation
field operators for an electron, and ${\bf r}$ and ${\bf r}'$ are two points along the edge.
In the limit of large $|{\bf r}$-${\bf r}'|$, the Green's function behaves as 
\begin{equation}
G_{edge}({\bf r})\sim |{\bf r}$-${\bf r}'|^{-\alpha}
\end{equation} 
which defines the edge exponent $\alpha$.  
The wave function of the state $\Psi_e({\bf r})|\psi>$ is proportional to  
$\psi({\bf r},{\bf r}_1,{\bf r}_2, ... {\bf r}_{N-1})$, obtained from the ground state wave function by
the replacement of one of the particle coordinates, say ${\bf r}_N$, by ${\bf r}$.  
The equal time Green's function can therefore be written as:
\begin{equation}
G_{edge}(|{\bf r}-{\bf r}'|)= N \frac{\int \prod_{j=1}^{N-1} d^2{\bf r}_j 
\psi^*({\bf r},\{{\bf r}_j\})\psi({\bf r}',\{{\bf r}_j\})} {\int 
\prod_{k=1}^{N} d^2{\bf r}_k \psi^*(\{{\bf r}_k\})\psi(\{{\bf r}_k\})}
\end{equation}
which can be evaluated efficiently by Monte Carlo for any given wave function for the ground state.

The composite fermion theory of the FQHE \cite{Jain} provides  
the wave function 
\begin{equation}
\psi^{(0)}_{\frac{n}{2n+1}}=P_{LLL}\prod_{j<k}(z_j-z_k)^2\phi_n
\end{equation}
for the FQHE state at $\nu=n/(2n+1)$.
Here $z_j=x_j-iy_j$ denotes the position of the $j$th particle, $\phi_n$ is the Slater determinant
wave function for $n$ filled Landau levels, and $P_{LLL}$ projects out the part of the wave function
that has component residing outside the lowest electronic Landau level. 
The factor $\prod_{j<k}(z_j-z_k)^2$ attaches two vortices to each electron in $\phi_n$; the 
bound state comprised of an electron and two quantized vortices is interpreted as a particle, called
the composite fermion, and the wave function $\psi^{(0)}_{\frac{n}{2n+1}}$ is interpreted as $n$ filled
Landau levels of composite fermions.  The microscopic wave functions and their interpretation in
terms of composite fermions have both been established.

\begin{figure}
\vspace{-2.5cm}
\centerline{\psfig{figure=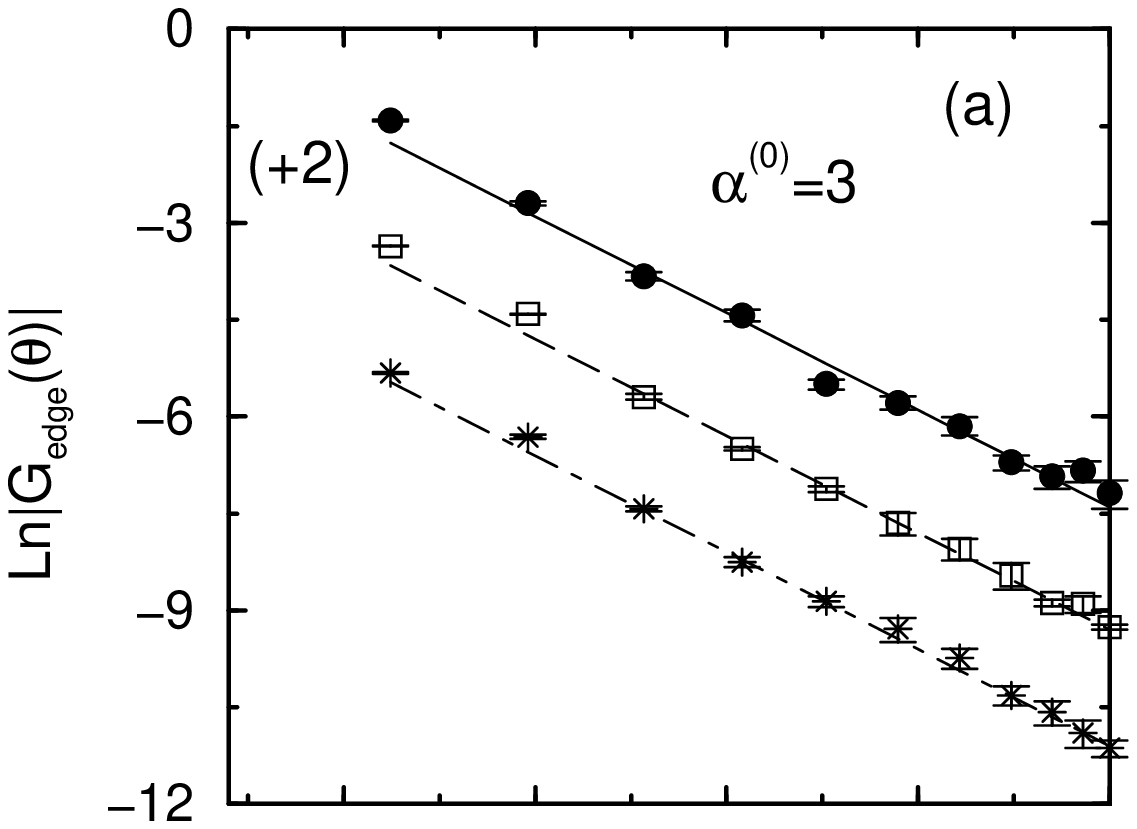,height=4.0in,angle=-90}}
\vspace{-6.0cm}
\centerline{\psfig{figure=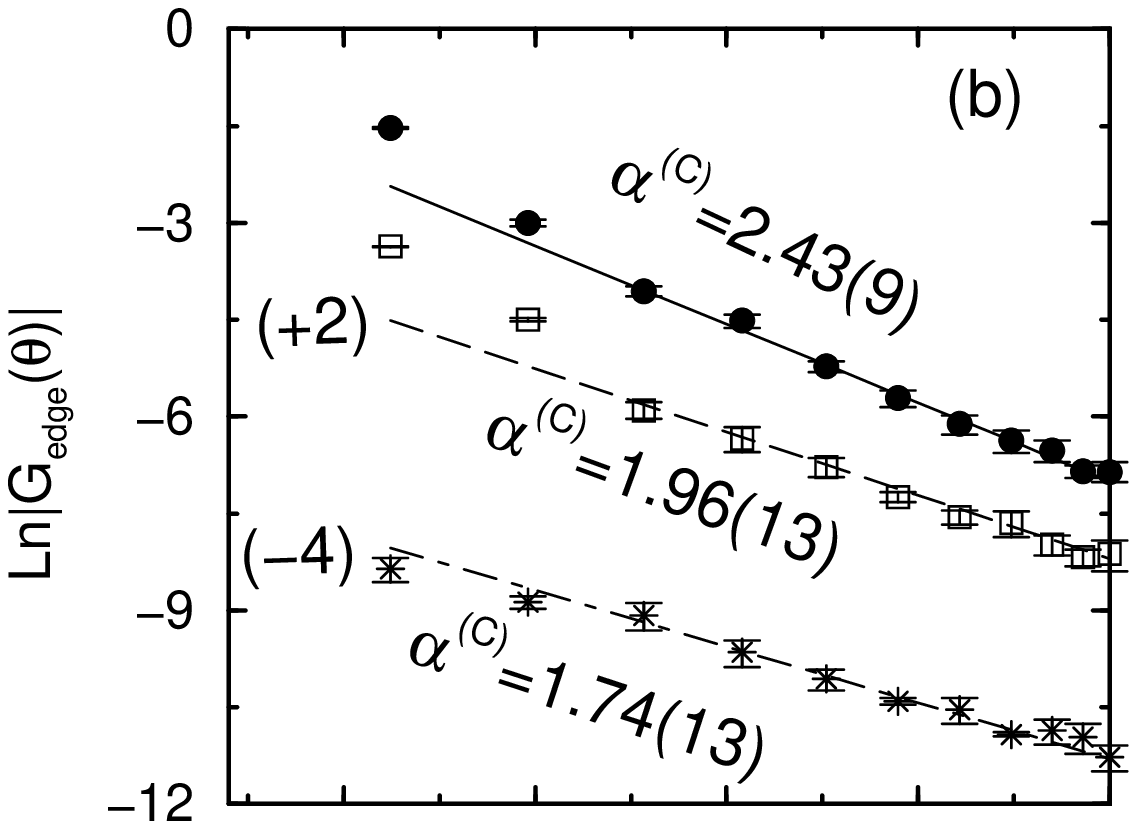,height=4.0in,angle=-90}}
\vspace{-6.0cm}
\centerline{\psfig{figure=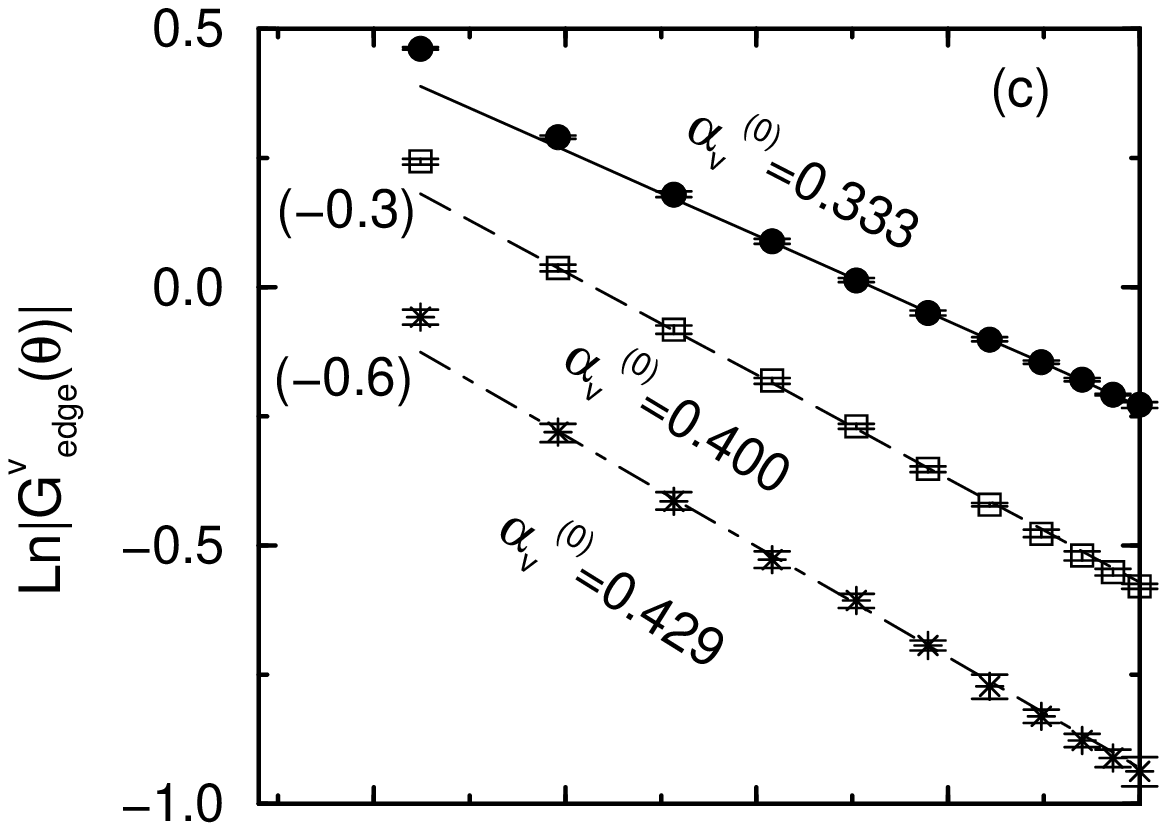,height=4.0in,angle=-90}}
\vspace{-6.0cm}
\centerline{\psfig{figure=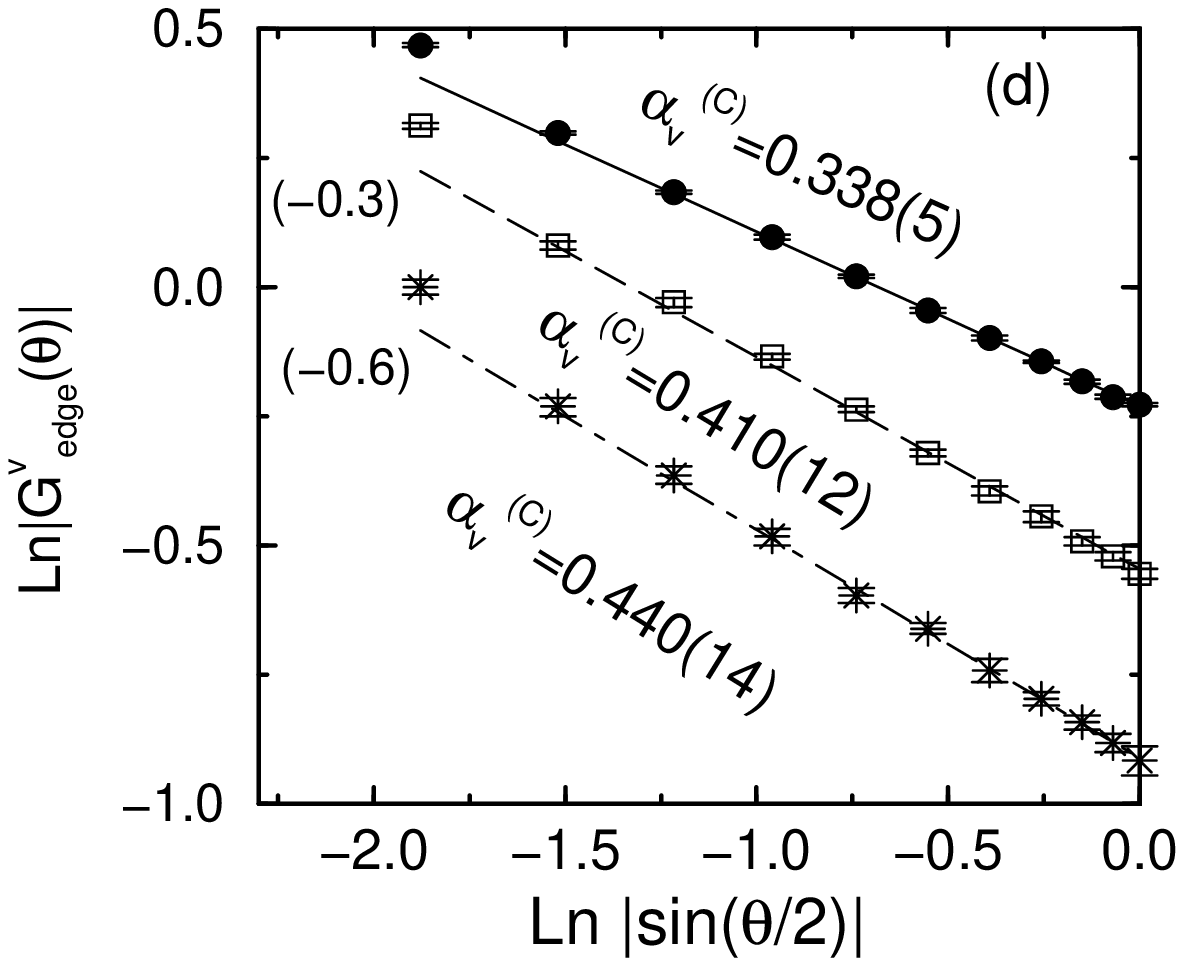,height=4.0in,angle=-90}}
\vspace{-2.5cm}
\caption{The correlation function $G_{edge}(|{\bf r}-{\bf r}'|) \equiv G_{edge}(\sin\theta/2)$ is
plotted as a function of $\sin(\theta/2)$ for 1/3, 2/5, and 3/7 (from
top to bottom, respectively, in each panel).  The points ${\bf r}$ and ${\bf r}'$ are chosen 
at the edge of the disk, at a distance of $R=\sqrt{2N/\nu}\l$ from the center,
where $l$ is the magnetic length,  and $\theta$ is the angle between ${\bf r}$ and ${\bf r}'$. 
The error bars indicate the statistical uncertainty in Monte Carlo.
The exponent $\alpha$ defined by $G(\theta) \sim |\sin(\theta/2)|^{-\alpha}$ 
is shown on the figure for each case.  For clarity, some lines have been shifted vertically by 
an amount given in parentheses on the left.  The panel (a) gives the exponent for
non-interacting composite fermions ($\psi^{(0)}$), panel (b) for interacting 
composite fermions ($\psi^{(C)}$), and panels (c) and (d) contain the vortex
correlation function, defined in text, for non-interacting and interacting composite fermions.
Systems with $N=40$, 50 and 60 composite fermions are used for 1/3, 2/5, and 3/7, respectively.}
\label{fig1}
\end{figure}

The calculations are performed for composite fermions 
confined inside a disk \cite{Dev,JK}.  There is a slight ambiguity regarding  
which state corresponds to $\nu=n/(2n+1)$ for $n>1$.  For example, for $N=30$ composite fermions at
$\nu=2/5$, we could take the configuration $(15,15)$, $(14,16)$, or $(16,14)$, where $(N_0,N_1)$
refers to the state containing $N_0$ composite fermions in the lowest composite-fermion Landau level
and $N_1$ in the  second.  Fortunately, we have found that the Green's functions for 
these choices differ only at short distances but not in exponent describing the long-distance behavior.
Therefore, we confine our attention to states that have equal numbers of composite fermions in each
composite-fermion Landau level. 
We have considered fully polarized states at $\nu=1/3$, 2/5, and 3/7, with the 
lowest Landau level projection evaluated in the standard manner \cite{JK}.  Confinement to a disk is
achieved by fixing the total angular momentum, which corresponds to a parabolic confinement
potential.  The calculated Green's functions, shown in Fig.~(\ref{fig1}a), 
are consistent with $\alpha^{(0)}=3$.  This was known for 1/3,  but is non-trivial for 
2/5 and 3/7, for which the wave function $\psi^{(0)}$ is 
rather complex.  The prediction from the effective theory thus correctly describes the edges 
of $\psi^{(0)}$.

The wave functions $\psi^{(0)}_{\frac{n}{2n+1}}$ describe {\em non}-interacting composite fermions, 
because $\phi_n$ is the ground state of non-interacting electrons.  These are known 
to be excellent approximations for the actual ground states of interacting electrons\cite{JK}, but 
they are not exact; the interaction between composite fermions is weak but finite, and  
leads to slight corrections to $\psi^{(0)}$.  This is of no consequence to the quantization 
of the Hall resistance, which remains unaffected so long as there is a gap in the excitation spectrum;
that is why it is often valid to neglect the CF-CF interaction in that context.
We now ask if that is also the case for the edge physics.

The effect of interaction between composite fermions is to cause mixing with higher 
{\em CF}-LLs.  (This ought to be distinguished from mixing with higher 
{\em electronic} LLs, which is neglected throughout this work.)
To incorporate the effect of CF-LL mixing, we 
diagonalize the Coulomb Hamiltonian in the basis 
$(\psi^{(0)}, \{\psi^{(0)p-h}\})$, where $\{\psi^{(0)p-h}\}$ denote states 
containing a single particle-hole pair of composite fermions \cite{Mandal}, and 
can be constructed explicitly from the corresponding electronic wave functions at 
filling factor $n$.  Various inner products required for an orthonormalization of the basis 
as well as the Coulomb matrix elements are evaluated by Monte Carlo \cite{Mandal}.  The ground state
thus obtained is denoted $\psi^{(C)}$ and the corresponding exponent $\alpha^{(C)}$.

As seen in Fig.~(\ref{fig1}b), $\alpha^{(C)}$ is significantly smaller than $\alpha^{(0)}=3$.
The calculations are performed for finite systems, containing up to 40, 50, and 60 particles 
for 1/3, 2/5, and 3/7, and the possibility that the exponent may change on the way to 
the thermodynamic limit cannot be ruled out in principle, but  
several facts suggest that our study captures the asymptotic physics:
The maximum distance along the edge is 30 times the characteristic length, namely the 
magnetic length; the system is big enough to produce a well 
defined exponent; the ``expected" exponent is obtained for $\psi^{(0)}$; 
and finally,  increasing the number of particles from 30 to 50 for 2/5 and 30 to 60 for 3/7
does not appreciably alter the exponent, while going from 30 to 40 particles at 
1/3 {\em reduces} $\alpha^{(C)}$ slightly \cite{Mandal}.

\begin{figure}
\vspace{-2.0cm}
\centerline{\psfig{figure=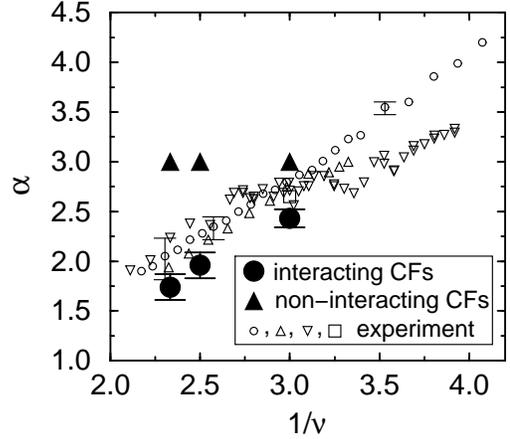,height=5.0in,angle=-90}}
\vspace{-1.5cm}
\caption{Tomonaga-Luttinger exponent, $\alpha$, for the FQHE edge liquid as a function of the filling
factor, $\nu$.  The filled circles (filled triangles) show theoretical values  
for interacting (non-interacting) composite fermions at $\nu=1/3$, 2/5, and 3/7.  The 
error bars refer to the statistical uncertainty coming from Monte Carlo as well as 
the linear fitting in Fig.~(\protect\ref{fig1}b).  
The experimental results (empty symbols) are taken from the following sources: square from Chang 
{\em et al.} \protect\cite{Chang1}; 
circles and triangles from Grayson {\em et al.}\protect\cite{Grayson} (samples M and Q); 
inverted triangles from Chang {\em et al.}\protect\cite{Chang3} (samples 1 and 2).}
\label{fig2}
\end{figure}

Fig.~(\ref{fig2}) shows a comparison between our theory and experiment.
The theoretical results for interacting composite fermions capture 
the qualitative behavior seen in experiment.  The systematic quantitative discrepancy between theory 
and experiment can be  ascribed to the neglect, in our calculation, 
of certain experimental features that could provide
corrections, for example disorder, the actual form of the confinement potential, or the 
screening of the interaction by the nearby gate.
We note that the tunneling experiments probe the {\em time} dependence 
of the Green's function, with the relevant correlation function 
being $G_{edge}({\bf r},t; {\bf r},0)$; however, for TL liquids  
it is expected that the behavior along the time direction is also described by the same exponent.

To gain insight into how the CF-CF interactions might enter into the edge physics, let 
us recall some facts about the TL approach to a one dimensional system of {\em chiral} 
fermions.\cite{TL} Given the commutator for the density operator:
\begin{equation}
[\rho_{-q'},\rho_q]=\frac{qL}{2\pi}\delta_{qq'}
\end{equation}
where $L$ is the length and $q$ the wave vector, one defines 
\begin{equation}
a_q=-i\sqrt{\frac{2\pi}{qL}}\; \rho_{-q}, \;\; a_q^{\dagger}=i\sqrt{\frac{2\pi}{qL}}\; \rho_{q}
\end{equation}
which satisfy $[a_{q'},a^{\dagger}_q]=\delta_{qq'}$.  One then defines the 
bosonic field 
\begin{equation}
\Phi(x)=\sum_{q>0} \sqrt{\frac{2\pi}{qL}}(e^{-iqx} a_q + e^{iqx} a^{\dagger}_q) e^{-a|q|/2}
\end{equation}
where $a$ is a regularization cut-off, to be set to zero at the end. 
The electron field operator can be written as 
$\Psi_e(x) \sim e^{-i\Phi(x)}$, an 
identity that can be rigorously established at the operator level.

We consider below $\nu=1/m$, where $m=2p+1$ is an odd integer. 
Wen argued that for the FQHE edge problem 
\begin{equation}
[\rho_{-q'},\rho_q]=\frac{1}{m}\frac{qL}{2\pi}\delta_{qq'}
\end{equation}
so the operators $a$ and $a^\dagger$ acquire a factor of $\sqrt{m}$ 
\begin{equation}
a_q=-i\sqrt{m}\sqrt{\frac{2\pi}{qL}}\; \rho_{-q}, \;\; a_q^{\dagger}=i\sqrt{m}\sqrt{\frac{2\pi}{qL}}
\;\rho_{q}
\end{equation}
The bosonic field operator $\Phi(x)$ 
is defined as above in terms of the new creation and annihilation operators. 
A key step in Wen's theory is the postulate that the electron field operator is given by 
\begin{equation}
\Psi_e(x) \sim e^{-i\sqrt{m}\Phi(x)}
\label{wen}
\end{equation}
This identification is consistent with antisymmetry
\begin{equation}
\{\Psi_e(x),\Psi_e(x')\}=0
\label{anti}
\end{equation}
and can also be shown to create an excitation with unit charge:
\begin{equation}
[\rho(x),\Psi_e^{\dagger}(x')]=\delta(x-x')\Psi_e^{\dagger}(x')
\label{charge}
\end{equation}
Various correlation functions can be evaluated straightforwardly with the help of Eq.~(\ref{wen}).

Eq.~(\ref{wen}) can be justified microscopically \cite{Wen} for Laughlin's wave function \cite{Laughlin} 
\begin{equation}
\psi_{1/m}^{(0)}=\prod_{j<k}(z_j-z_k)^m \exp[-\frac{1}{4}\sum_i|z_i|^2]
\end{equation}
The vortex excitation at $\eta$ is given by $\prod_{j}(z_j-\eta)\psi_{1/m}^{(0)}$.
Wen showed, employing an analogy to a two-dimensional one-component classical plasma \cite{Laughlin},
that the vortex excitation at the edge of Laughlin's wave function is equivalent to
$e^{i\Phi(x)/\sqrt{m}}$.  The operator $e^{i\sqrt{m}\Phi(x)}$ creates $m$ vortices at $\eta$,
given by $\prod_{j}(z_j-\eta)^m\psi_{1/m}^{(0)}$.  However, this is precisely the wave 
function obtained by the
application of $\Psi_e(\eta)$ on the $N+1$ particle Laughlin's wave function.  The equivalence of a hole
and $m$ vortices establishes Wen's {\em ansatz} for Laughlin's wave function.

However, this derivation does not carry over to other possible wave functions at $\nu=1/m$.
The form of the general wave function at $1/m$ is $\prod_{j<k}(z_j-z_k)F[\{z_i\}]$, 
where $F[\{z_i\}]$ is a symmetric
function.  Creation of a hole at $\eta$ amounts to replacing $z_N\rightarrow \eta$, 
which produces $\prod_{j}(z_j-\eta)\prod_{j<k}(z_j-z_k)F[z_N=\eta,\{z_i\}]$.  This has a 
single order-one vortex at $\eta$.  Treating the wave function as a function of one 
of the coordinates, say $z_1$, we expect that the wave function typically has $m-1$ 
additional vortices near $\eta$, the exact positions of which depend on the
coordinates of the other particles.

For an arbitrary ground state $\psi$, we define a vortex 
\begin{equation}
\psi_V(\eta)=\prod_{j}(z_j-\eta)\psi
\end{equation}
Because $\prod_{j}(z_j-\eta)\phi_n$ has a hole in each Landau level, $\psi_V(\eta)$ describes
the state with a CF hole in each CF-Landau level.  In the
interior, the vortex has a charge equal to $\nu e$ relative to the neutral background, 
but its charge is not quantized near the edge.  
The vortex-vortex correlation function is defined as
\begin{equation}
G^V_{edge}(|\eta-\eta'|)\sim \frac{\int d^2{\bf r}_1...d^2{\bf r}_N 
\psi_V^*(\eta')\psi_V(\eta)} {\int d 
^2{\bf r}_1...d^2{\bf r}_N |\psi|^2}
\end{equation}
The plots in Fig.~(\ref{fig1}c,d) indicate that $G^V$ also has a power law behavior, 
governed by an exponent $\alpha^{(0)}_V\approx \alpha^{(C)}_V\approx \nu$ 
that is independent of the actual ground state, suggesting 
that the vortex excitation is to be identified with the vertex operator $e^{-i\sqrt{\nu}\Phi(\eta)}$.
This assignment implies that the analogous Green's function for a
multiple vortex, created by multiplication by $\prod_{j}(z_j-\eta)^n$ has an exponent $n^2\nu$ associated
with it; we have confirmed that as well.  These results are in line with the predictions of 
Ref.~\onlinecite{Wen}.

Thus, it appears that while the vortex excitation has a simple representation in the effective
one-dimensional theory, the electron field operator $\Psi_e(x)$ does not.
Before concluding, we speculate on the possibility that $\Psi_e(x)$ might be represented by 
a non-local operator in the one-dimensional problem.  This should not 
be surprising because quite often, especially for non-trivial mappings,
simple, local operators of one theory are mapped into complicated, non-local ones in the new  
theory.  Let us consider 
\begin{equation}
\Psi_e(x)\equiv \int dy g(|y-x|) e^{i\sqrt{m}\Phi(y)}
\end{equation}
where $g(|y-x|)$ is a normalizable function peaked at $y=x$.  
Eq.~(\ref{anti}) implies antisymmetry: $\Psi_e(x)\Psi_e(x')=-\Psi_e(x')\Psi_e(x)$, and 
Eq.~(\ref{charge}) shows that $\Psi_e(x)$ creates an excitation of charge one:
\begin{equation} 
[\hat{N},\Psi_e^{\dagger}(x')]=\int dx [\rho(x),\Psi_e^{\dagger}(x')]=\Psi_e^{\dagger}(x')
\end{equation}
But now the equal time Green's function is given by
\begin{equation}
<\Psi^{\dagger}_e(x) \Psi_e(x')> \sim \int dy \int dy' \frac{g(|y-x|) g(|y'-x'|)}{ (y-y')^{m}}  
\end{equation}
If $g(|x-y|)$ has a finite range, then a quantized exponent is obtained as before.
On the other hand, if one assumes a power-law form $g(|x-y|)\sim |x-y|^{-\beta}$, 
then from dimensional considerations, we get  
\begin{equation}
|<\Psi^{\dagger}_e(x) \Psi_e(x')>| \sim |x-x'|^{-\alpha} 
\end{equation}
with $\alpha=m-2(1-\beta)$.
The normalizability of $g(x)$ requires $\beta>1/2$, and the requirement that the above integrals be well
defined at coincident points 
imposes the condition $\beta<1$.  Together, these imply that $\alpha$ lies between $m$ and $m-1$.
While the above discussion is only speculative, showing that, at least in principle, 
non-locality can lead to a non-quantized exponent, it is worth noting 
that all theoretical and experimental exponents for $\nu=1/3$ lie between 2 and 3.

It is a pleasure to acknowledge partial support by the National Science Foundation 
under Grant No. DMR-9986806.  
We are grateful to A. Chang, G. Murthy, and A. Sen for illuminating discussions and 
A. Chang for freely sharing his data with us.

\pagebreak

\pagebreak

\begin{references}
\bibitem{shoulder} There is sometimes a shoulder observed in $\alpha$ at $\alpha\approx 2.7$ 
that is identified with $\nu=1/3$; the fact that it is not centered at $\nu=1/3$ is thought 
to be related to the difference between edge and bulk densities \protect\cite{Chang3}.
\end{references}
\end{document}